\documentclass[conference]{IEEEtran}
\IEEEoverridecommandlockouts
\usepackage{booktabs}

\usepackage{cite}
\usepackage{comment}
\usepackage{amsmath,amssymb,amsfonts}
\usepackage{algorithmic}
\usepackage{graphicx}
\usepackage{multirow}
\usepackage{textcomp}
\usepackage{xcolor}
\usepackage{array} 
\usepackage{amssymb}  

\def\BibTeX{{\rm B\kern-.05em{\sc i\kern-.025em b}\kern-.08em
    T\kern-.1667em\lower.7ex\hbox{E}\kern-.125emX}}
\begin{document}

\title{Transforming Decoder-Only Transformers for Accurate WiFi-Telemetry Based Indoor Localization}
\author{
    \IEEEauthorblockN{Nayan Sanjay Bhatia, Katia Obraczka} 
     
    \IEEEauthorblockA{University of California, Santa Cruz \\Santa Cruz, California, USA \\ \{nbhatia3, katia\}@ucsc.edu}
}

\maketitle

\begin{abstract}
Wireless Fidelity (WiFi) based indoor positioning is a widely
researched area for determining the position of devices within a wireless network. Accurate indoor location has numerous applications,
such as asset tracking and indoor navigation. Despite advances in WiFi localization techniques- in particular approaches that leverage WiFi telemetry- their adoption in practice remains limited due to several factors including environmental changes that cause signal fading, multipath effects, interference, which, in turn, impact positioning accuracy. In addition, telemetry data differs depending on the WiFi device vendor, offering distinct features and formats; use case requirements can also vary widely. Currently, there is no unified model to handle all these variations effectively. In this paper, we present WiFiGPT, a Generative Pretrained Transformer (GPT) based system  
that is able to handle these variations while achieving high localization accuracy. Our experiments with WiFiGPT demonstrate that GPTs, in particular Large Language Models (LLMs), can effectively capture subtle spatial patterns in noisy wireless telemetry, making them reliable regressors. Compared to existing state-of-the-art methods, our method matches and often surpasses conventional approaches for multiple types of telemetry. Achieving sub-meter accuracy for RSSI and FTM and centimeter-level precision for CSI demonstrates the potential of LLM-based localisation to outperform specialized techniques, all without handcrafted signal processing or calibration.
\end{abstract}

\section{Introduction}
Indoor positioning technologies have various applications, ranging from indoor navigation in environments such as airports, hospitals, and shopping malls, accessibility, health- and elderly care to improving smart building  systems~\cite{9531633,farahsari2022survey,correa2017review}. However, its adoption in the real world has been slow due to practical issues. Many localization systems are trained and tested on static environments in a controlled manner. However, the real world is dynamic, and even slight environmental variations can cause catastrophic loss in accuracy~\cite{dwiyasa2016survey}. In order to make real impact, localization approaches must be able to adapt to different dynamic scenarios while maintaining high accuracy~\cite{deak2012survey}. 
More recently, localization using WiFi telemetry has been receiving considerable attention from researchers and practitioners due to WiFi's ubiquity, cost-effectiveness and scalability~\cite{roy2022survey}. WiFi telemetry offers valuable information about the Radio Frequency (RF) environment through metrics such as the Received Signal Strength Indicator (RSSI), Channel State Information (CSI) and Fine Timing Measurement (FTM)~\cite{10609490,8692423}. This telemetry, especially the CSI, characterizes how a signal propagates from a transmitter to a receiver, encompassing effects like scattering, fading, and power decay~\cite{cominelli2023exposing}.
As AI and ML tools become increasingly prevalent, they have been used for indoor localization systems to improve accuracy, adaptability, and real-time performance in a variety of environments~\cite{nessa2020survey}. Indoor localization has traditionally relied on rule-based algorithms that need telemetry features like RSSI, CSI, or FTM to be manually customized. While effective in custom settings, these models often fail to generalize across environments and require significant engineering effort to adapt to different target scenarios. Common neural networks or non-linear regressor based approaches often require data preprocessing steps such as filtering and data scaling. These methods rely on fixed schemas and struggle with missing or incomplete data, limiting flexibility and scalability across heterogeneous environments~\cite{twala2009empirical}.

As Transformers~\cite{NIPS2017_3f5ee243} become more ubiquitous and accessible, we have explored their use to address the challenges posed by accurate WiFi-based indoor localization.
Transformers~\cite{NIPS2017_3f5ee243} were originally built for language translation, where the goal is to understand textual sequences and generalize across different contexts. Their generalization capability makes Transformers a good candidate for modeling WiFi telemetry, which is often dynamic, noisy, sparse, and affected by signal propagation impairments. In environments that are heavily affected by the effect of multipath, like corridors or indoor spaces, even small changes in position can lead to significant signal variations, which traditional methods struggle to capture. Transformers, and more broadly LLMs, provide a flexible way to learn patterns from spatio-temporal data without needing hand-crafted features. Their generalization ability, which has already been shown in NLP, computer vision, robotics, and even biomedical signals~\cite{pires2019multilingual,google2024dolphingemma,liu2025can,10.1145/3677846.3677854}, can be applied here to make sense of complex patterns in WiFi environments. In other words, instead of building different models for every setup, we can train a single model that learns to work across environments.

Motivated by the cross-generalization capabilities of GPTs, in particular LLMs, we investigate their potential to address indoor localization tasks using WiFi signals to accurately estimate positioning in dynamic environments subject to noise and multipath effects. 
In this paper, we introduce WiFiGPT, a decoder-only large language model adapted and re-tooled to model real-world WiFi telemetry for accurate indoor localization. LLMs have been typically trained to model textual sequences in natural language. In WiFiGPT, we leverage decoder-only LLMs' learning capability to model the distribution of telemetry signals, treating each signal feature as an input and mapping the final answer to the device position. We demonstrate that LLMs are not limited to text-based systems and can operate in a noisy environment utilizing wireless telemetry data for indoor positioning. The system is designed with portability, agility, and adaptability in mind. It can quickly adjust to different environments while maintaining high accuracy. This allows fast learning, even from just a few samples, and enables strong cross-generalization. Because of this, the resulting model can be rapidly trained and deployed in real-world scenarios with minimal computational overhead where LLMs already exist, such as smart home devices, smartphones.
\begin{table*}[!t]
\caption{Comparison of Wireless Localization Techniques}
\label{table:localization_comparison}
\centering
\begin{tabular}{|l|p{2.3cm}|p{2cm}|p{2cm}|p{1.6cm}|}
\hline
\textbf{Feature} & \textbf{CSI} & \textbf{RSSI} & \textbf{Time-Based} & \textbf{AoA} \\
\hline
\textbf{Accuracy} & High & Low to Moderate & Moderate to High & High \\
\hline
\textbf{Resolution} & Fine-grained & Coarse-grained & Fine-grained & Fine-grained \\
\hline
\textbf{H/W Requirements} & Specialized (e.g., Intel 5300 NIC) & Standard WiFi &  Timing hardware & Multiple antennas \\
\hline
\textbf{Computational Complexity} & Moderate to High & Low & Moderate & High \\
\hline
\textbf{Multipath Resilience} & Good (can leverage multi-path) & Poor & Moderate & Good \\
\hline
\textbf{Ease of Implementation} & Moderate to Complex & Simple & Moderate & Complex \\
\hline
\end{tabular}
\end{table*}%
In this paper, we make the following contributions:
\begin{itemize}
    \item To the best of our knowledge, WiFiGPT is the first system repurposing decoder-only LLMs for indoor localization. 
    \item We present a generalized framework with deterministic outputs that integrates diverse telemetry sources and adapts across radio environments. Extending LLMs beyond language, we apply a simple training scheme to map telemetry sequences to distance predictions for sensor-based regression.
    \item We show that LLMs can handle WiFi telemetry directly without manual feature engineering. WiFiGPT performs well on real-world indoor datasets and adapts quickly, even with limited training data, getting sub-meter accuracy.
    \item We analyze how the accuracy of different LLaMA model sizes varies under changing environmental conditions.
    \end{itemize}

The rest of the paper is organized as follow: 
In Section II, we provide an overview of the background and related work setting the stage for our contributions. 
Section III describes in detail our large WiFi-telemetry model for indoor localization, while in Section IV, we present its implementation details. Section V describes how we evaluated the performance of our system and Section VI presents and discusses our results. Finally, Section VII concludes the paper and discusses directions for future work. 

\section{BACKGROUND AND RELATED WORK} 

In this section, we start with a brief overview of the background on WiFi telemetry and then describe the state-of-the-art approaches including both traditional models and more recent methods based on large language models (LLMs). We highlight the strengths and limitations of each, setting the stage for our proposed approach.
\subsection{WiFi Telemetry}
WiFi provides different telemetry data, such as Received Signal Strength Indicator (RSSI), Channel State Information (CSI), Angle of Arrival (AoA) or time-based systems like Fine Timing Measurement (FTM), Time of Flight (ToF), Return Time of Flight (RToF) each offering trade-offs between accuracy, complexity, and practicality for Indoor Positioning. (Table I). 

\noindent
{\bf RSSI:}
Received Signal Strength Indicator (RSSI) is a metric used to quantify the strength of a received signal during wireless communication~\cite{pagano2015indoor}. 
RSSI-based techniques are commonly used because they are easily accessible and involve minimal overhead~\cite{mistry2015rssi}. RSSI values are often scaled based on the specific hardware and manufacturer specifications, meaning there is no standardized scale to compare data. Fingerprinting approaches~\cite{yiu2017wireless} are common. However, it is sensitive to environmental noise and device heterogeneity. Hence, even after enhancement of ML and sensor fusion~\cite{singh2021machine}, RSSI-based approach often shows high localization error. To reduce the error, RSSI is used in conjunction with other telemetry data. The process of combining telemetry data from different sensors is called Sensor Fusion. In 2021, Microsoft's indoor localization competition by Microsoft~\cite{lymberopoulos2017microsoft} attracted participation from 1,170 global teams exploring various methodologies. The dataset encompasses dense signatures of WiFi, GMF, iBeacons, BLE and ground truth (waypoint) locations from numerous buildings in Chinese cities, and the top teams achieved high localization accuracy using the sensor fusion setup using non-NN models like KNN and LGBM with an error margin of 1.3 m.

\noindent
{\bf FTM:}
IEEE 802.11-2016~\cite{9733026} included the first generation of the FTM protocol that sends a burst of frames and averages the round-trip-time (RTT). In the Line-Of-Sight (LOS) environment, FTM can perform significantly better than RSSI, with errors around two meters~\cite{ibrahim2018verification}. However, not all commercial devices support it; accuracy is affected by clock skew and drift and needs calibration accordingly. Its performance suffers in Non-Line-Of-Sight environments (NLOS), where positioning error exacerbates and can be as high as 5-10 meters in NLOS settings~\cite{bullmann2020comparison,ibrahim2018verification}.

\noindent
{\bf CSI and AoA:}
CSI captures environmental conditions by providing the
phase and amplitude of different subcarriers. Consequently, the system is more robust against scattering, fading and power delays. Indoor localization has seen advancements in the Non-Line-of-Sight (NLOS) environment through systems like Ubicarse, ArrayTrack, and SpotFi\cite{kumar2014accurate,xiong2013arraytrack,10.1145/2785956.2787487}, each based on using Angle of Arrival(AoA) derived from the Channel State Information(CSI). MUSIC (Multiple Signal Classification)~\cite{qian2017enabling} and ESPRIT (Estimation of Signal Parameters via Rotational Invariance Techniques)~\cite{paulraj1985estimation} signal processing technique that separates signal and noise spaces and can be used to estimate angles of arrival (AoA) of signals in array signal processing. All AoA methodology are based on some variation of MUSIC and ESPRIT requires multiple antennae to get phase differences. Hence, most existing work relies on specialized hardware as traditional routers have firmware restrictions and cannot work on single-antenna systems (ESP32, Raspberry Pi)~\cite{espressif_esp32,raspberrypi_foundation}. This makes them less ubiquitous than RSSI and FTM. System performance is also affected by feedback delays, and errors introduced by the use of outdated or incorrect CSI. 
Acquiring accurate CSI introduces requires control message exchanges between the transmitter and receiver, restricting valuable bandwidth by adding overhead.

\subsection{LLMs and Their Applications}
Transformer models have demonstrated remarkable adaptability across diverse domains. In particular, as shown in ~\cite{pires2019multilingual}, large language models (LLMs), which can be realized by generative pre-trained transformers (GPTs), which are trained primarily in English, still perform well in other languages. Since then, transformers' capacity for generalization has expanded far beyond natural language processing to include applications such as computer vision (vision transformers), robotics, structured code generation, modeling vital signs, e.g., eletrocardiogram (ECG) signals, and even interpreting dolphin vocalizations~\cite{google2024dolphingemma,liu2025can,10.1145/3677846.3677854}. 

Transformers can be broadly classified into two types: Models that can handle all tokens at once (encoders) and those that can only handle past tokens (decoders)~\cite{qorib2024decoder}.
Encoder-based transformers such as BERT~\cite{koroteev2021bert} take full input and process it simultaneously. This requires access to the whole input sequence before any processing can begin (non-causal). However, they cannot be used when data is available in real time (such as WiFi packets) and are more suitable for offline tasks such as recovering lost network packets~\cite{zhao2024mininglimiteddatasufficiently} or data imputation~\cite{cesar2023bert}. 
Decoder-based transformers generate an output sequence that is not dependent on the future element (causal). Decoder-only transformers are also called autoregressive, in that each output is generated one step at a time based on current and past elements. Decoder-only architectures are thus well suited for real-time applications. Contemporary systems, such as Openai's GPT~\cite{ai2023gpt} and Meta's LLaMA~\cite{touvron2023llama}, adopt a decoder-only architecture. Throughout the paper, we use the term Generated Pretrained Transformers (GPT) for decoder-based transformers only.

LLMs are gaining traction in networking applications. NetLLM~\cite{Wu_2024} introduces a framework that adapts large language models (LLMs) for networking tasks, including viewport prediction, adaptive bitrate streaming, and cluster job scheduling. WirelessLLM~\cite{shao2024wirelessllmempoweringlargelanguage} deals with challenges like power allocation and spectrum sensing in communication networks and SigLLM~\cite{alnegheimish2024largelanguagemodelszeroshot} explores the application of LLMs for time series anomaly detection. CSI-BERT~\cite {Zhao_2024} uses an encoder-based transformer to recover lost CSI in wireless sensing applications caused by packet loss. LocGPT~\cite{10.1145/3643832.3661869} proposes a custom encoder-decoder transformer architecture for antenna array-based triangulation using phase spectra. It requires extensive pre-training data to learn general localization features.

Vacareanu et al. (2024)~\cite{vacareanu2024words} show that LLMs can perform regression purely through in-context learning, achieving competitive results on nonlinear synthetic datasets without fine-tuning. Tang et al. (2024)~\cite{tang2024understanding} show that LLM embeddings can be used for regression tasks in higher-dimensional settings than traditional feature engineering on synthetic datasets. 
Building on this, we show that exising decoder-only LLMs can accurately perform distance calculations. Once fine-tuned on a sufficient set of WiFi packets, our system can deliver accurate distance predictions without requiring large-scale pretraining or extensive scene-specific data.





\begin{figure*}[ht]
    \centering
    \includegraphics[width=0.7\textwidth]{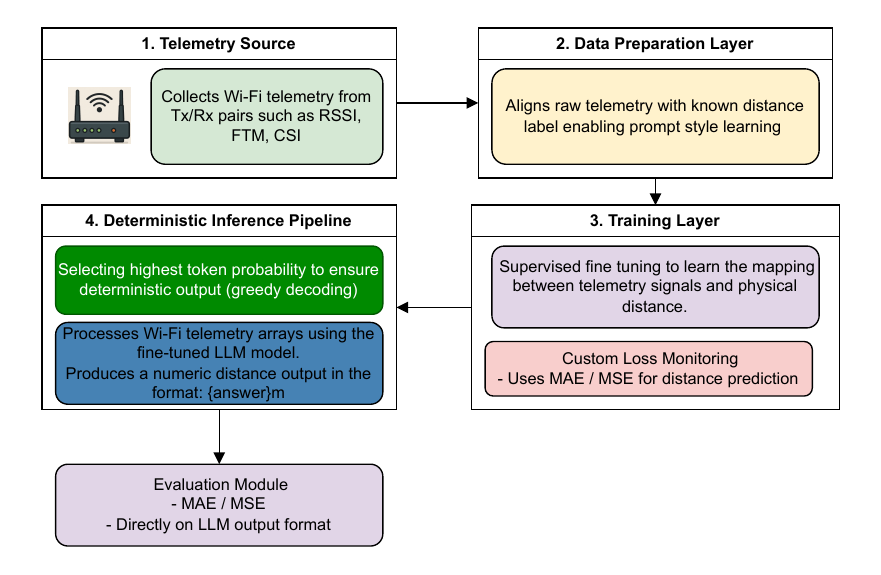}
    \caption{WiFiGPT: System Flow}
    \label{fig:locagpt_arch}
\end{figure*}
\section{WiFiGPT}

WiFiGPT (Figure 1) is a deterministic language model-based system designed for indoor distance estimation using Wi-Fi telemetry. It begins by collecting telemetry data from transmitter-receiver (Tx/Rx) pairs. This raw data is then aligned with ground-truth distance labels to enable prompt-style learning. In the training phase, a large language model is fine-tuned in a supervised manner to map telemetry sequences to corresponding distances, using MAE and MSE as custom loss metrics. During inference, the model applies greedy decoding to output a single numeric distance prediction in meters, ensuring deterministic behavior. The system is evaluated by computing MAE/MSE directly on the output format, making it a streamlined and interpretable solution for localization tasks.

\subsection{Telemetry Source}
We use telemetry data captured between transmitters (Tx) and receivers (Rx) of various WiFi devices, including ESP32 devices, Google Access Points and Pixel smartphones. Specifically, our system shows results for multisource telemetry inputs such as CSI, FTM, and RSSI measurements. However, our model is not restricted to these particular types of telemetry and can handle other telemetry and sensor data. Our method processes these telemetry signals sequentially through an autoregressive model, allowing the system to generalize distance predictions across different environments and hardware setups. Importantly, by learning directly from raw telemetry data, our model remains strong and works well with any device. It adapts reliably to different indoor conditions without relying on specific hardware or signal types.
\subsection{Data Preparation Layer}
A token is a chunk of character sets that the model reads or generates step by step. Large Language models are primarily autoregressive models that means they do next token prediction based on the previous inputs and is expressed by the equation: 
\begin{equation}
P(x_1, x_2, \ldots, x_T) = \prod_{t=1}^{T} P(x_t \mid x_1, \ldots, x_{t-1})
\label{eq:autoregressive}
\end{equation}
where \( x_t \) is the token at position \( t \), predicted based on previous \(x_1... x_{t-1} \) tokens. 

To adapt the LLM for a regression-style task using sensor telemetry, we present the problem as a language modelling goal, where the model is familiarized with the telemetry input. The model learns to generate the target value as the next token in a structured prompt-response format. By framing the positioning task as a next-word prediction problem, we avoid modifying the LLM architecture, eliminating the need of a encoder for input data. The output token must be a numeric value in a consistent and parseable format. It should be deterministic and agnostic to the system conditions. 
The telemetry data needs to be represented effectively in a structured format so that LLM can understand the effective relationship and predict the assign task. We modify equation (1) as follows:
\begin{equation}
 P(\texttt{\{Distance\}m} \mid \text{Instruction}, \text{[Telemetry]})
\end{equation}
We treat the instruction prompt and telemetry data as the input tokens $x_1, x_2, \ldots, x_{t-1}$ and the position as token $x_t$. Reframes the LLM modeling objective into a regression task.
The LLM is trained using a JSONL-based instruction tuning format~\cite{agarwal2025think} where each WiFi packet is structured as a JSON object containing a single text field. We use four delimiter tokens for different sections of our prompts. \texttt{<\textbar begin\_of\_text\textbar>} and \texttt{<\textbar end\_of\_text\textbar>} indicate the start and end of each new packet, respectively. \texttt{<\textbar start\_header\_id\textbar>} and \texttt{<\textbar end\_header\_id\textbar>} creates a distinct identity between the user and the assistant. The user agent can send available WiFi telemetry or other sensor data (BLE, gyroscope), and the assistant returns the estimated distance in the format: \{answer\}m. The \texttt{<\textbar end\_of\_text\textbar>} also acts as end of sequence (EOS) token. 
eg. 
\begin{quote}
\small
\texttt{<|begin\_of\_text|><|start\_header\_id|> user <|end\_header\_id|>}\\
\texttt{based on the array predict distance in meter and nothing else in this format: \{answer\}m.}\\
\texttt{this is array: [WiFi Telemetry]}\\
\texttt{<|start\_header\_id|> assistant <|end\_header\_id|> \{Distance\}m <|end\_of\_text|>}
\end{quote}
This format is used for alignment \cite{shen2023large}. Formatting the model response in this way simplifies automated evaluation in later phases. During inference, the text until the end header ID is provided to the LLM, prompting it to predict the distance as the next token in this format \{answer\}m.

Unlike traditional deep learning methods like CNNs, LSTMs, or nonlinear regressors like LGBM and KNN, the LLM-based approach does not require data scaling or converting categorical features (like vendor information or room numbers) into one-hot vectors. A fixed schema is not required, and missing features can be omitted. It can convert wireless environment information into high-dimensional vectors, capturing semantic meanings and relationships into the latent space.  For instance, the user commands, "Turn on the nearest light." The system can calculate the user's precise location, understands the contextual intent, and activates the nearest light. The model could answer questions like "Where is the nearest restroom?" or "What is the signal strength near Room 404?". 
\subsection{Training}




\subsubsection{Select Base LLM Model}
We select the Meta's LLaMA 3 family of models, including variants 1B~\cite{mlxcommunity2024llama32}, 3B~\cite{mlxcommunity2024llama32_3b}, and 8B~\cite{mlxcommunity2025deephermes3}, due to their open-source availability, strong performance characteristics, and established presence in the research literature as representative examples of modern decoder-only transformer architectures. LLaMA has open‑weights, permissive open-source license with strong benchmark scores for language tasks and a longer context window size, allowing us to process a whole packet in one pass~\cite{zhao2024llama}.
\subsubsection{Fine Tuning}
Large Language Models (LLMs) can model a task when given examples of that task in their context, exhibiting Few-Shot and Zero-Shot Capabilities. We conduct supervised fine-tuning (SFT)~\cite{dong2023abilities} using LoRA (Low-Rank Adaptation) adapters~\cite{hu2022lora} on the LLaMA models. We used Apple's MLX framework~\cite{mlx2023} on a 32GB M1 Max chip. SFT involves training the LLM model on the positioning-labeled data, while LoRA enables efficient fine-tuning of only selected model layers instead of the whole network. LoRA is an efficient
way to adapt base language models to new tasks without
changing the entire layer, which can be both time- and computing-intensive. LoRA freezes the original weights and updates only a fraction of the parameters. After training,
it creates a LoRA adapter for the given dataset to be
loaded/unloaded for different tasks. Reduces the computational overhead of training on consumer-grade hardware, allows modularity, and allows us to easily swap or stack adapters for different tasks without retraining the entire model. We use batch size 2, number of layers 16, and learning rate (lr) 2e-4. We use a smaller batch size and num-layers due to limited GPU memory, which can lead to noisier gradients~\cite{zhang2024scaling}. We use the relatively standard learning rate to fine-tune~\cite{unsloth_finetuning_guide}.
We train the LLM model separately for each dataset and environment to evaluate how it responds to different signal distributions and spatial layouts. This allows us to study its behavior in a controlled setting and to understand how well it captures environment-specific patterns. According to the Chinchilla Law~\cite{hoffmann2022trainingcomputeoptimallargelanguage}, models benefit more from increased training data than from an increased parameter count. Therefore, as we accumulate more radio environment data across diverse settings, the model should generalize better in future deployments, even without retraining, by leveraging its prior exposure to similar propagation characteristics.
\subsection{Deterministic Inference}
LLMs are generative models that learn statistical patterns in training data to predict the next token in a sequence. As a result, they are inherently stochastic, meaning that their output can vary depending on the decoding strategy used. We used deterministic text generation (greedy decoding) by setting the temperature t to 0.0, top-p to 1.0 and do-sample to false. We also set a fixed random seed to ensure that the results are fully reproducible every time we run the experiments. The temperature parameter controls randomness in the generation process by scaling the logits (raw output scores produced by the model) before applying the softmax function. Temperature 0.0 ensures that the model consistently selects the most probable token. Top-p value of 1.0 enables consideration of the full token distribution. Setting do-sample to false forces the model to not do sampling and ensures greedy decoding. The random seed establishes a fixed initialization point for any probabilistic components, guaranteeing identical outputs given the same input conditions.

\begin{itemize}
    \item \textbf{Temperature} \( T = 0.0 \) scales the logits (\(z\)) such that \( P(x) = \text{softmax}(z/T) \rightarrow \arg\max \) as \( T \rightarrow 0 \), forcing deterministic selection.
    \item \textbf{do-sample} \( = False \) restricts candidate tokens to the single most likely token.
    \item \textbf{Top-p} \( = 1.0 \) includes the entire token distribution (no truncation by cumulative probability).
    \item \textbf{Random seed} fixes any stochastic elements (e.g., dropout, sampling noise), ensuring reproducibility.
\end{itemize}

Under these conditions, the output is fully deterministic; the same input prompt always allows the same output token.


 

\section{Experimental Methodology}

\subsection{Datasets}
We demonstrate the accuracy of our system using two datasets. The first dataset is custom collected, where we capture Channel State Information (CSI) using two ESP32-S2 devices positioned 6 to 10 meters apart in a hallway with human presence, recording data for around 5 minutes and maintaining LOS. The second data set is publicly available~\cite{zenedo}, combining WiFi fine-time measurement (FTM) and received signal strength indicator (RSSI) data collected in three environments: a lecture theater (LOS), an office (mixed LOS-NLOS), and a corridor (NLOS) to improve positioning accuracy under varying conditions. Using these two datasets underscores our model's adaptability and robustness across telemetry and environmental scenarios.
\subsubsection{ESP-32-CSI-HALLWAY}
Hallways are a multipath-prevalent environment~\cite{9155240}.
ESP32 is a single antenna device; therefore, we cannot extract phase difference information between antennas. Therefore, we cannot use state-of-the-art algorithms like MUSIC because they are inapplicable for similar low-cost devices. This limitation also applies to many low-cost Android phones and even the Raspberry Pi, making the single-antenna constraint particularly relevant in such scenarios. 

We use two ESP32-S2 devices operating at 20 MHz, one as a transmitter and the other as a receiver. The ESP32 chip we are using uses 802.11n WiFi 4. 802.11n was the first generation to introduce CSI.
The devices are placed 6 to 10 meters apart at 1-meter intervals, recording CSI for approximately 5 minutes in a hallway. The environment includes multipath noise and people occasionally lingering around, but the setup maintains line-of-sight (LOS).
The data set is custom collected to ensure that Meta's LLaMA model is not exposed to these data, since the specifics of LLaMa's training data are not publicly available. The data set is structured as follows:
\begin{quote}
\small
\texttt{<|begin\_of\_text|><|start\_header\_id|> user <|end\_header\_id|>}\\
\texttt{based on the array predict distance in meter and nothing else in this format: \{answer\}m.}\\
\texttt{this is array: [84 -64 4 0 0 0 0 0 0 0 0 0 13 1 14 2 14 3 14 4 14 3 13 4 12 3 13 4 13 4 11 4 11 5 10 5 9 5 9 5 8 5 7 5 6 4 5 4 4 4 3 3 3 3 2 2 1 2 1 2 0 2 -1 2 0 0 -3 1 -2 2 -4 2 -5 2 -5 2 -6 3 -6 4 -7 4 -7 5 -7 3 -9 5 -9 5 -9 6 -10 7 -11 7 -11 8 -10 8 -10 9 -10 10 -9 11 -10 12 -9 13 -9 15 -9 16 -11 16 -10 17 0 0 0 0 0 0 0 0 0 0 ]}\\
\texttt{<|start\_header\_id|> assistant <|end\_header\_id|> \textbf{6m} <|end\_of\_text|>}
\end{quote}

\subsubsection{WiFi RSS and RTT dataset with different LOS conditions for indoor positioning}
\begin{table}[ht]
\centering
\caption{WiFi RSS \& RTT dataset with different LOS conditions for indoor positioning\cite{feng2024wifi}}

\begin{tabular}{|l|c|c|c|}
\hline
\textbf{Data features} & \textbf{Lecture Theatre} & \textbf{Office} & \textbf{Corridor} \\
\hline
Testbed (m$^2$) & 15 $\times$ 14.5 & 18 $\times$ 5.5 & 35 $\times$ 6 \\
Grid size (m$^2$) & 0.6 $\times$ 0.6 & 0.6 $\times$ 0.6 & 0.6 $\times$ 0.6 \\
Number of RPs & 120 & 108 & 114 \\
All samples & 7,200 & 6,480 & 6,840 \\
Training samples & 5,280 & 4,860 & 5,110 \\
Testing samples & 1,920 & 1,620 & 1,740 \\
WiFi condition & LOS & LOS-NLOS & NLOS \\
\hline
\end{tabular}
\end{table}

We use a publicly available WiFi dataset that comprises various LOS and NLOS environments~\cite{zenedo}. It contains an extensive selection of samples from multiple reference points (RP) in three different scenarios: lecture theater (LOS), corridor (mixed LOS / NLOS) and office (NLOS) (Table II). Data are collected using a Google AC-1304 WiFi router and Pixel 3 smartphones. The data set contains RTT and RSSI measurements, as both the Access Point (AP) and the Station (STA) support the FTM protocol. 
The dataset is structured as follows:
\begin{quote}
\small
\texttt{<|begin\_of\_text|><|start\_header\_id|> user <|end\_header\_id|>}\\
\texttt{based on the array predict distance in meter and nothing else in this format: \{answer\}m.}\\
\texttt{this is array: [WiFi FTM AP 1 (ns): 11351.0, WiFi RSSI AP 1 (dbm): -63.0, WiFi FTM AP 2 (ns): 5599.0, WiFi RSSI AP 2 (dbm): -60.0, WiFi FTM AP 3 (ns): 2918.0, WiFi RSSI AP 3 (dbm): -52.0, WiFi FTM AP 4 (ns): 13573.0, WiFi RSSI AP 4 (dbm): -68.0, WiFi FTM AP 5 (ns): 10671.0, WiFi RSSI AP 5 (dbm): -64.0]}\\
\texttt{<|start\_header\_id|> assistant <|end\_header\_id|> \ \textbf{18.25m} <|end\_of\_text|>}
\end{quote}

\subsection{Evaluation Metrics}
Training large-language models with MSE or MAE instead of cross-entropy loss can create fundamental issues. MAE (Mean Absolute Error) and MSE (Mean Squared Error) loss functions assume continuous, symmetric error distributions, incompatible with the probabilistic nature of next-token prediction that cross-entropy handles. We are interested in monitoring the training progress without modifying the transformer architecture. Therefore, we insert the loss functions as an auxiliary module to observe whether the model converges. 
\begin{equation}
\text{MSE} = \frac{1}{n} \sum_{i=1}^{n} (y_i - \hat{y}_i)^2
\label{eq:MSE}
\end{equation}
\begin{equation}
\text{MAE} = \frac{1}{n} \sum_{i=1}^{n} |y_i - \hat{y}_i|
\label{eq:MAE}
\end{equation}

\noindent where:
\begin{itemize}
    \item $y_i$ represents the actual value
    \item $\hat{y}_i$ represents the predicted value
    \item $n$ is the total number of samples
\end{itemize}
These evaluation metrics are auxiliary diagnostic tools; we do not use them for gradient updates during training. This allows us to evaluate the regression behaviour of the LLM model while preserving the underlying cross-entropy-based language modelling intent. We keep a misalignment counter additionally to track outputs that are misaligned or cannot be parsed. These are skipped in the loss metrics and logged separately for quality checks allowing us to monitor whether the LLM is learning to align with the regression task.
\section{Results}
\begin{table}[h!]
\centering
\caption{Performance of LLaMA-3 Models on 51,500 CSI Samples (80\% of the Dataset)}
\setlength{\tabcolsep}{4pt}      
\begin{tabular}{ccccc}
\toprule
\textbf{Model} & \textbf{MSE (m)} & \textbf{MAE (m)} & \textbf{R\textsuperscript{2}} & \textbf{Misalignments} \\ \\
\midrule
LLaMA-3.2-1B & 0.1809 & 0.0759 & 0.9163 & 0 \\
LLaMA-3.2-3B & 0.1678 & 0.0688 & 0.9224 & 6 \\
LLaMA-3.1-8B  & 0.3772 & 0.1809 & 0.8255 & 2 \\
\bottomrule
\end{tabular}
\end{table}
\begin{table*}[ht]
\centering
\caption{Model performance on Testing samples across different environments}
\begin{tabular}{|c|c|r|r|r|r|r|r|r|r|}
\hline
\textbf{Parameter} & \textbf{Environment} & \textbf{MSE} & \textbf{MAE} & \textbf{R\textsuperscript{2}} & \textbf{25th} & \textbf{50th} & \textbf{75th} & \textbf{100th} & \textbf{Iteration} \\
\hline
1b & Corridor & 3.0567  & 1.3359 & 0.9891  & \textbf{0.9600}  & 1.0000  & 2.0000  & 10.0100  & 2750 \\
3b & Corridor & \textbf{1.8553} & \textbf{1.1166}  & \textbf{0.9934 }& 0.9750 & 1.0000 & \textbf{1.0000} & \textbf{9.98}& 4000 \\
8b & Corridor & 2.4009 & 1.2155 & 0.9914 & 0.9800 & 1.0000 & 1.9900 & 10.0500 & 4350 \\
\hline
1b & Theatre & 1.8281 & 1.0328 & 0.9557 & 0.47 & 0.83 & 1.36 & 9.34 & 2550 \\
3b & Theatre & \textbf{1.23085} & \textbf{0.90263} & \textbf{0.97} & 0.4299 & 0.8299 & 1.08 & \textbf{5.37} & 2500 \\
8b & Theatre & 1.5360 & 0.9601 & 0.9628 & 0.4900 & 0.76 & 1.36 & 6.71 & 3750 \\
\hline
1b & Office & 2.4462 & 1.1589 & 0.9641 & 0.39 & 0.8 & 1.85 & 5.98 & 2150 \\
3b & Office & 2.2595 & 1.0974 & 0.9668 & \textbf{0.29} & 0.9 & 1.58 & 7.22 & 3500 \\
8b & Office & \textbf{1.9816} & 1.107094 & \textbf{0.97094} & 0.47 & 0.97 & 1.76 & \textbf{5.37} & 2400 \\
\hline
\end{tabular}
\end{table*}
\subsection{Few Shot Learning-CSI}
To evaluate how the model behaves with a few shots of learning, we split the CSI dataset by class, assigning 10\% for training, 10\% for validation, and the remaining 80\% for testing. We chose this configuration to present a realistic few-shot learning scenario. 
Table III presents the performance of various LLaMA models on 51,500 CSI samples collected in a hallway environment with strong multipath propagation and signal reflections from the surrounding, including human presence. Despite these challenging conditions, the CSI-based models overall achieved near-perfect accuracy. The LLaMA-3.2-3B model led in performance, reaching the lowest MSE of 0.1678 m and MAE of 0.0688 m, although it experienced six misalignments. The LLaMA-3.2-1B model had slightly higher error metrics (MSE: 0.1809 m, MAE: 0.0759 m, $R^2$: 0.9163) but confirmed perfectly stable predictions (0 misalignments). LLaMA-3.1-8B model performed slightly worse on all metrics, although not significantly. These results highlight the effectiveness of LLMs fine-tuned with CSI data in enabling accurate indoor localization, even in complex multipath indoor environments.
\begin{figure}[h!]
    \centering
    \includegraphics[width=\linewidth]{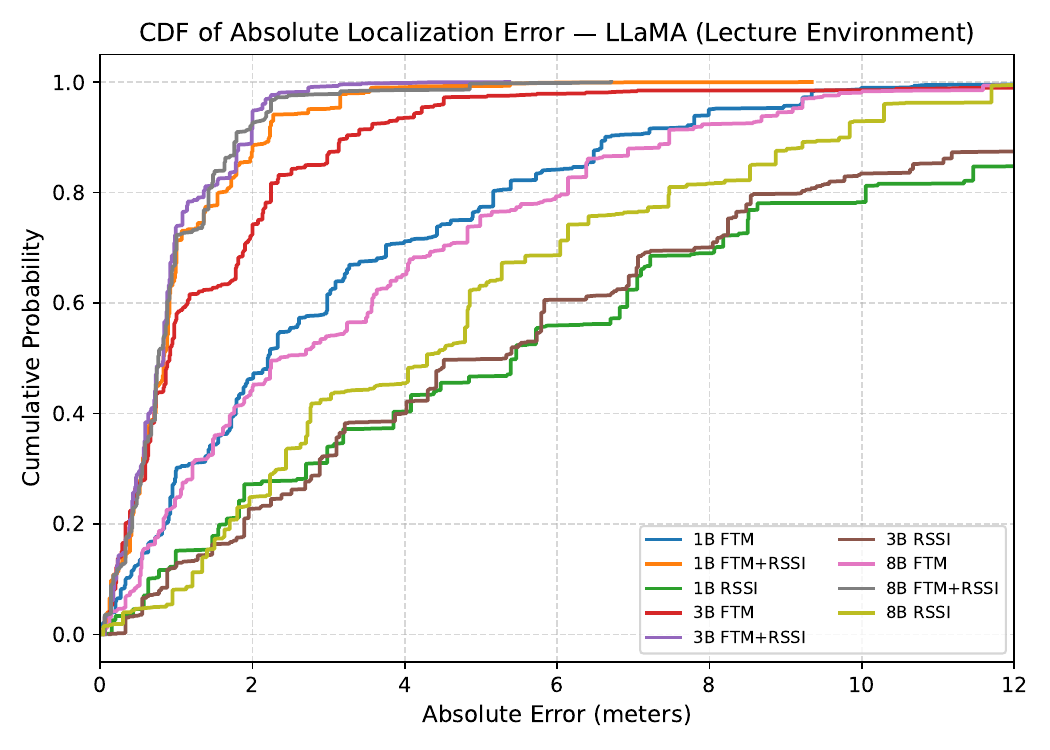}
    \caption{CDF of Localization Error}
    \label{fig:localization_cdf}
\end{figure}

\subsection{FTM and RSSI}
The results (Table IV) show a clear trend in which the environmental conditions (LOS and NLOS) and the complexity of the model affect the positioning accuracy. Across all setups, LLMs exhibit strong regression capabilities, with median localization errors typically near or below 1 meter. 

In the fully NLOS Corridor environment, the 3B model has the highest accuracy, with an MAE of 1.12 m, and MSE of 1.86. The smaller 1B model also remains effective in such scenario with the lowest 25th percentile (0.96) error showing its capability in constrained, noisy environment.  

Theater (LOS) has the best performance with the 3 billion model with the lowest MSE (1.23) suggesting few outliers and lowest MAE (0.90) suggesting consistent performance overall. The LOS condition and relatively square shape (15 × 14.5 $m^2$) likely contributing to the higher accuracy. The theater has the highest number of RPs (120), which also explains better spatial granularity and improved results.

In contrast, in the office environment, which presents mixed LOS-NLOS conditions, the 8B model outperforms others with an MSE of 1.98 highlighting its ability to generalize better in heterogeneous signal environments. Nevertheless, the 1B model notably achieves the lowest MAE (approximately 1.15 m) and the smallest errors at the median (0.81 m) making it suitable for resource-constrained deployment scenarios.

Across all environment, LLMs exhibit strong regression capabilities, with median localization errors typically near or below 1 meter. These findings demonstrate the practical use of decoder-only architectures to integrate diverse WiFi telemetry data (RSSI, FTM, CSI) within a unified learning framework, enhancing indoor localization without relying on explicit rule based models. This sets the system accuracy suitable for indoor navigation, asset tracking, and augmented reality (AR) use cases.

\subsection{Data Imputation}
 Large Language Models (LLMs) often perform implicit data imputation to compensate for missing input features. By framing the positioning problem through a language-based interface, we are able to leverage this capability to enhance robustness and generalization. In the background and related work, we've already mentioned that combining multiple telemetry sources (sensor fusion) improves accuracy, especially since FTM, a time-based mechanism, offers finer granularity than coarse-grained mechanisms like RSSI. To test whether our model has learned to use these modalities dynamically, we use a model trained on both FTM and RSSI data and then selectively ablate one feature set at a time. 
 Figure 2 shows the cumulative distributed probability of model performance based on the choice of telemetry selected.
\subsubsection{FTM+RSSI} 
Across all model sizes (1B, 3B, 8B), the FTM+RSSI combinations consistently achieve the highest cumulative probability at lower error thresholds. This supports the theoretical premise that fusing time-based (FTM) and power-based (RSSI) features provides complementary information, improving generalization and implying that the model can capture the intricacies.
\subsubsection{FTM-Only Ablation Study}
FTM performs the second best. The FTM-only models (especially 3B FTM and 8B FTM) show strong performance, closely following the FTM+RSSI curves. This aligns with the fact that time-based telemetry perform better than course grained RSSI. It is worth noting that the 1B model exhibited 296 misalignments, the 3B model reduced this to 22, and the 8B model achieved perfect alignment with zero misalignments out of 1920 samples used for testing.
\subsubsection{RSSI-Only Ablation Study}
RSSI-only models consistently lag behind in performance across all model sizes, with slower CDF growth and heavier tails. This is expected due to RSSI’s high sensitivity to multipath and environment-specific attenuation, which limits its localization precision. It is also worth noting that 1b had 8 misalignments whereas other models didn't have any misalignments.
\subsection{Discussion}

Our methodology analyzes the accuracy using a single WiFi packet from each AP to isolate models' behaviour. However, a single packet may be corrupted in practical deployments due to hardware imperfections, clock drift, or other system-level limitations beyond the model's control, implicating tail latency. To mitigate these issues, most systems rely on multiple packets. For instance, even with CSI (most precise form of WiFi telemetry) SpotFi~\cite{10.1145/2829988.2787487} requires 10 CSI packets to achieve a 0.5 median accuracy while worst error reaches up to 10m.

Practitioners adopting our system can apply a sliding window of 5–10 packets to smooth out tail-heavy errors, incurring only a few milliseconds of additional delay. Alternatively, the LLM can be adapted to ingest multiple packets jointly rather than processing each individually. However, it is important to note that increasing the input context window also increases the time complexity in a quadratic manner~\cite{ali2024tokenizer}. LLMs can also be given the physical dimensions of the environment in the prompt itself to prevent hallucinations beyond the defined boundary conditions.

Figure 2 showcases that even with incomplete telemetry (RSSI-only), larger models (e.g., 8B RSSI) still show improvement over smaller counterparts (e.g., 1B RSSI). This indicates that the LLaMA model effectively leverages its internal latent space, learning correlations and compensating for missing features during fine-tuning. The model’s ability to infer distance from imperfect data (e.g., RSSI-only cases) demonstrates that the vector embeddings capture semantically rich structure, allowing practical distance estimation even under partial observability. The bigger parameter models showcase no misalignments whereas the smaller model tends to misalign with ablated data. However, as noted in ~\cite{hoffmann2022trainingcomputeoptimallargelanguage}, model performance improves more with increased data than just scaling model size. Once the smaller models are trained on a bigger corpus of radio data, they can generalize further and achieve better accuracy. That’s why we trained the model separately for each environment and dataset, so we could closely observe how it behaves under different conditions without external data. As long as the model gets relevant data, it can generalize effectively.

Rather than using embeddings or few-shot prompts, we treat indoor positioning as a next-token prediction task, aligning the output distribution with predicted distance as the final token. This approach uses LLMs' autoregressive modelling, internal attention mechanisms, positional encodings, and emergent sequence forecasting behaviours, capturing sequential dependencies that static methods miss. Our simplified pipeline consists of a single autoregressive pass that is generalizable across sensors and environments.


\section{Limitations and Future work}
We selected the LLaMA family of models for our initial experiments because they are open source, available in different sizes (1B, 3B, 8B), and are widely adopted in the research community. Moving forward, we plan to explore other model families like Gemma and Phi, and closed-source LLMs such as Claude and OpenAI's ChatGPT. We also aim to study the effects of quantization across different models to understand better the trade-offs in performance, efficiency, and deployment feasibility.

LLMs are known to be sensitive to prompt phrasing due to their autoregressive property~\cite{anagnostidis2024susceptiblellmsinfluenceprompts}. Even a slight change of wording can drastically affect the next token, potentially hindering its ability to do regression. We plan to study the influence of prompts on indoor positioning accuracy. As a temporary workaround, we can treat the regression task as a separate model call using the structured query described in the training data format (as outlined in the paper). 
\begin{figure}[h!]
  \centering
  \includegraphics[width=0.5\textwidth]{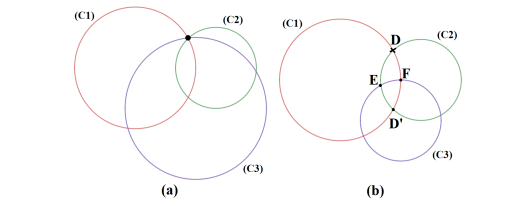}
  \caption{Trilateration geometry: (a) Ideal Scenario (b) Real Scenario~\cite{9287413}}
  \label{fig:trilateration}
\end{figure}

Our system's performance may degrade in a new environment without any prior training or Dataset of new sensors. If there is no error, the device should meet at a singular point in space as shown in Figure 2a~\cite{9287413}. However, errors due to signal fading, path loss, create a geometry instead of a centroid, as seen in Figure 2b. Calculating the area of the geometry and having multiple access points(more than 3) allows us to get the ground truth. Smaller error area (\(A\)) implies higher accuracy, so we define the feedback as \(f_{\text{feedback}} = \frac{1}{A}\). Without the ground truth, this feedback serves as a reward signal, minimizing localization errors.

Our approach shows great potential for localization accuracy. Even though our model runs on commodity hardware, it requires significant computing power. However, that is starting to change. There has been a large-scale effort to reduce computing costs while making models more power efficient~\cite{shekhar2024towards,ma2024era}. For example, Microsoft’s BitNet b1.58~\cite{ma2025bitnetb1582b4ttechnical} reduces memory and computing needs without losing much performance. With breakthroughs like this, more models are moving toward low-bit quantization. We can expect these models to run directly on consumer Stations (STA) or affordable Access Points (AP) that can support multiple user without costing a fortune.
\section{Conclusion}
In this paper, we introduce WiFiGPT, a telemetry-agnostic, token-based WiFi positioning system built entirely on large language models. Using a custom training template for the telemetry dataset and fine-tuning Llama 3 using LoRA, we could adapt a purely autoregressive decoder to a regression task. We show it can reliably capture complex signal behaviours like multipath without filtering or preprocessing while consistently staying under 1m error. The pipeline, which includes an auxiliary regression monitor and a fully deterministic inference setup, makes LLMs a reliable regressor. This setup shows that token-based regression is a viable alternative to traditional methods. It is schema-less by design, requires no preprocessing pipeline, and effectively captures semantic relationships for categorical data. It also highlights how large language models can implicitly learn nuanced spatial patterns in noisy wireless telemetry by attending to structured input sequences and understanding the language of wireless.

\bibliographystyle{IEEEtran}
\bibliography{references}

\vspace{12pt}

\end{document}